\documentclass[a4paper]{jpconf}
\usepackage{amsmath,amscd,amssymb}
\usepackage{graphicx}
\usepackage{cite}
\pdfoutput=1
\newcommand{\sfract}[2]{{\textstyle\sqrt{\frac{#1}{#2}}}}
\begin{document}
\title{Kink--Antikink Scattering in $\varphi^4$ and $\phi^6$  Models}

\author{Herbert Weigel}

\address{Physics Department, Stellenbosch University, Matiland 7602, South Africa}

\ead{weigel@sun.ac.za}

\begin{abstract}
For kink--antikink scattering within the $\varphi^4$ non--linear field 
theory in one space and one time dimension resonance type configurations 
emerge when the relative velocity between kink and antikink falls below a 
critical value. It has been conjectured that the vibrational excitation 
of the kink would be the source for these resonances because (simplified) 
collective coordinate calculations, that emphasized on this excitation, 
qualitatively reproduced those resonances. Surprisingly a numerical study 
in the $\phi^6$ field theory also exhibited such resonances even though 
it does not contain the vibrational excitation. To explore this 
contradiction we start from the working hypothesis that in
either model any collective coordinate ansatz which includes a degree of
freedom similar to the vibrational excitation leads to resonances, regardless
of whether or not this mode emerges as a solution to the (linearized) field 
equations.  To this end we compare numerical results in the $\varphi^4$ and
$\phi^6$ models that arise from the full set of partial differential equations 
to those from the ordinary differential equations for a collective coordinate
ansatz. An inaccuracy in literature formulas for the collective coordinate
approach in the $\varphi^4$ model requires to revisit those calculations.
\end{abstract}

\section{Motivation}

The kink soliton in the $\varphi^4$ model is often considered as the
prototype~\cite{Ra82} configuration for more elaborate soliton systems 
that occur in field theory. The range of physics disciplines in which 
solitons are relevant is huge. Solitons appear in 
cosmology~\cite{Vilenkin:1994,Vachaspati:2006}, particle and nuclear
physics~\cite{Manton:2004,Weigel:2008zz}, as well as condensed matter
physics~\cite{Bishop:1978}. Kink--antikink configurations are of 
particular interest because they can mimic particle--antiparticle
interactions and might eventually provide more inside into fundamental 
concepts like crossing--symmetry in particle scattering~\cite{Abdelhady}.
Using modern desktop computers, numerical solutions to the field equations
that initially represent a widely separated but approaching kink--antikink pair
are feasible. This amounts to numerically integrating partial differential 
equations (PDE) for time and space dependent fields. Nevertheless it is 
interesting to simplify those equations by assuming collective coordinate 
ans\"atze. This reduces the PDE to coupled ordinary differential equations 
(ODE) for a limited number of time dependent functions. Equally interesting, 
collective coordinates may also answer the question of which 
are important modes of the system. In this context, the so--called 
{\it shape mode} in the $\varphi^4$ model has attracted particular attention.
This mode is a bound state in the vibration spectrum about the kink~\cite{Ra82_5}.
It has been assumed to be responsible for the {\it bounce} type solutions 
to the PDE in kink--antikink interactions~\cite{Goodman:2005aa}. These bounce
solutions are resonating kink--antikink configurations. The perception is 
that the shape  mode can absorb sufficient energy from the kink--antikink 
system to prevent it from falling apart after collision. This role of the 
shape mode has been doubted as the $\phi^6$ model also contains bounce type 
solutions in the kink--antikink sector(s)~\cite{Dorey:2011yw}. However, 
apart from the translational zero mode there is no bound state in the 
vibration spectrum of the kink in the $\phi^6$ model. In this model thus 
a thorough investigation of the collective coordinate analysis is required 
to better understand this contradiction. We will do so in sections four and 
five and adopt the working hypothesis that a collective coordinate ansatz 
with a shape mode type component will be a suitable approximation in either 
case. Furthermore it has turned out that the collective coordinate calculations 
in the $\varphi^4$ model have inherited (typographical) errors in the formula 
for the source term of the shape mode from ref.~\cite{Sugiyama:1979mi}. We 
will therefore revisit those calculations in section three. In all cases 
we will compare results from the corresponding ODE to solutions of the 
PDE, the full field equations. The latter are obtained with programming 
code adapted from that used in ref.~\cite{Abdelhady:2011tm}.  We conclude 
and comment on the relevance of collective coordinate calculations in 
section six. We will, however, commence with a brief review of kink--antikink 
approaches.

\section{Kink--Antikink Concept}

\begin{figure}[t]
\centerline{
\includegraphics[width=10cm,height=5.5cm]{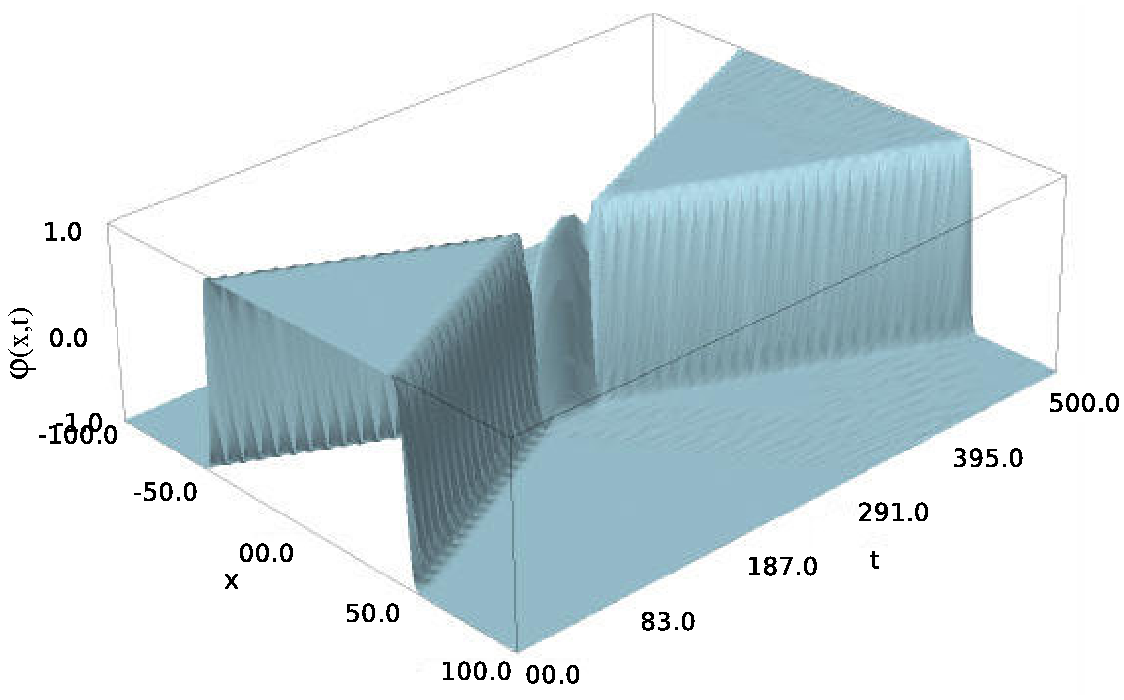}}
\caption{\label{fig:bounce}A typical solution to the PDE with 
kink--antikink initial conditions. Figure adopted from ref.~\cite{Abdelhady}.}
\end{figure}

By now there have been many detailed numerical studies
\cite{Kudryavtsev:1975dj,Campbell:1983xu,Belova:1985fg,Anninos:1991un,Goodman:2005aa} 
of the PDE for the field in the $\varphi^4$ model in one time
and one space dimension\footnote{See ref.~\cite{Goodman:2005aa} for a 
comprehensive summary of these computations.}. The model is defined 
by the Lagrangian
\begin{equation}
\mathcal{L}_4=\frac{1}{2}\partial_\mu\varphi\partial^\mu\varphi
-\frac{1}{2}\left(\varphi^2-1\right)^2\,,
\label{eq:lag1}
\end{equation}
where all model parameters have been absorbed by appropriate redefinitions
of the coordinates $(t,x)$ and the field $\varphi$. The field equation
reads
\begin{equation}
\ddot{\varphi}-\varphi^{\prime\prime}=2\varphi\left(1-\varphi^2\right)\,,
\label{eq:eom1}
\end{equation}
where dots and primes denote time and coordinate derivatives,
respectively. The (anti)kink $\varphi_{K,\overline{K}}=\pm{\rm tanh}(x)$ 
is a static solution to this equation. It connects the two vacuum 
solutions at $\varphi_{\rm vac}=\pm1$. Kink--antikink configurations
\begin{equation}
\varphi_{K\overline{K}}(x,X(t))
=\varphi_{K}(\xi_{+})+\varphi_{\overline{K}}(\xi_{-})-1
={\rm tanh}(\xi_{+})-{\rm tanh}(\xi_{-})-1
\label{eq:kkbar4}
\end{equation}
with
\begin{equation}
\xi_{\pm}=\frac{x}{\sqrt{1-v_{\rm in}^2}}\,\pm X(t)
\qquad {\rm and} \qquad
\dot{X}(0)=\frac{-v_{\rm in}}{\sqrt{1-v_{\rm in}^2}}
\label{eq:initial}
\end{equation}
are solutions for wide separation $X(t)\gg0$. For these conventions
$X(t)$ essentially measures the position of the antikink. More 
interestingly these configurations may serve as initial conditions 
for the equation of motion by choosing $X(0)$ large enough to avoid 
interference and the constant velocity $v_{\rm in}$ such that kink 
and antikink approach each other. A typical solution
is shown in figure~\ref{fig:bounce}.
The most interesting feature is the appearance of bounces for 
initial velocity $v_{\rm in}$ below a critical value $v_{\rm cr}$.
The example in figure~\ref{fig:bounce} has two bounces but 
solutions with many bounces (and traps) exist too~\cite{Goodman:2005aa}.
The occurrence of such bounces has frequently been linked to 
the existence of the so--called shape mode
\begin{equation}
\chi(x,t)={\rm e}^{i\omega_1 t} \chi_1(x)
\qquad {\rm with} \qquad
\chi_1(x)=\frac{{\rm sinh}(x)}{{\rm cosh}^2(x)}
\label{eq:shape}
\end{equation}
with eigen--frequency $\omega_1^2=3$ in the vibration 
spectrum of the kink~\cite{Ra82_5}. (The continuum spectrum starts
at $\omega^2=4$ for the present units.) The common argument is that 
the collective coordinate parameterization
\begin{equation}
\varphi_{\rm cc}(x,t)=\varphi_{K\overline{K}}(x,X(t))+
\sfract{3}{2}\,A(t)\left[\chi_1(x+X(t))-\chi_1(x-X(t))\right]\,,
\label{eq:ansatz}
\end{equation}
that reduces the PDE~(\ref{eq:eom1}) to the (simpler) ODE for the time 
dependent collective coordinates $A(t)$ and $X(t)$, not only approximates 
the full solution reasonably well, but also reproduces $v_{\rm cr}$
within 10\%~\cite{Goodman:2005aa}\footnote{Another argument
is based on the perturbed sine--Gordon model~\cite{Goodman:2004aa}.}.
In the above ansatz the difference of the shape modes at 
$\xi_\pm$ is assumed because the (anti)kinks do not directly excite the sum.

Similar bounces and traps have been recently observed~\cite{Dorey:2011yw}
in the $\phi^6$ model which is defined by the Lagrangian
\begin{equation}
\mathcal{L}_6=\frac{1}{2}\partial_\mu\phi\partial^\mu\phi
-\frac{1}{2}\phi^2\left(\phi^2-1\right)^2\,.
\label{eq:lag2}
\end{equation}
Although the corresponding field equation
\begin{equation}
\ddot{\phi}-\phi^{\prime\prime}=-\phi\left(3\phi^4-4\phi^2+1\right)\,,
\label{eq:eom2}
\end{equation}
also allows for (anti)kink solutions
$\phi_{K,\overline{K}}=\left[1+{\rm exp}(\pm2x)\right]^{-\frac{1}{2}}$,
there is no bound state with non--zero energy in the vibration 
spectrum about the kink. This suggests that the existence of the 
shape mode does not serve as a rigorous criterion for the 
occurrence of bounces and traps. 

In view of the above observed puzzle we reverse that line of 
argument into a working hypothesis. We want to test the assumption 
that the solutions of the ODE for the collective coordinate ans\"atze 
reasonably well approximate the solutions to the full PDE; regardless of 
whether or not $\chi(x,t)$ solves the field equation for vibrations in 
the kink background. For the $\varphi^4$ model the ansatz is given in 
eq.~(\ref{eq:ansatz}). There are two analogs in the $\phi^6$ model 
\begin{eqnarray}
\phi_{\rm cc}(x,t)&=&
\phi_{K}(\xi_{+})+\phi_{\overline{K}}(\xi_{-})
+\sfract{3}{2}\,A(t)\left[\chi_1(x+X(t))-\chi_1(x-X(t))\right]\,,
\label{eq:kbark} \\
\overline{\phi}_{\rm cc}(x,t)&=&
\phi_{K}(\xi_{-})+\phi_{\overline{K}}(\xi_{+})-1
+\sfract{3}{2}\,A(t)\left[\chi_1(x+X(t))-\chi_1(x-X(t))\right]\,,
\label{eq:kkbar}
\end{eqnarray}
since there are three possible vacuum solutions, $\phi_{\rm vac}=0,\pm1$.
We substitute these ans\"atze into the Lagrangian and integrate over the 
spatial degree of freedom. Formally the resulting Lagrangian for the 
collective coordinates reads
\begin{eqnarray}
L_6(A,\dot{A},X,\dot{X})&=&\int dx\, \mathcal{L}_6(\varphi_{\rm cc})\cr\cr
&=&a_1\dot{X}^2-a_2+a_3\dot{A}^2-a_4A^2+a_5A+a_6\dot{X}^2A+a_7\dot{X}\dot{A}\cr\cr
&& \hspace{0.4mm}
+a_8\dot{X}^2A^2+a_9A\dot{X}\dot{A}-a_{10}A^3-a_{11}A^4-a_{12}A^5-a_{13}A^6\,.
\label{eq:lagf}
\end{eqnarray}
The coefficient functions depend on the (relative) distance parameter, 
{\it i.e.} $a_i=a_i(X)$ for $i=1,\ldots,13$. They also have a parametrical 
dependence on the initial velocity $v_{\rm in }$ via the construction in 
eq.~(\ref{eq:initial}). The $\varphi^4$ model has a similar collective 
coordinate Lagrangian, just that $a_{12}$ and $a_{13}$ are absent. 
The actual form of the $a_i$, of course, depends on whether 
eq.~(\ref{eq:kbark}), eq.~(\ref{eq:kkbar}) or the $\varphi^4$ model is 
considered. As an example, we list
\begin{equation}
a_1(X)=\frac{1}{2}\int_{-\infty}^\infty dx
\left[{\rm e}^{2\xi_{+}}\left(1+{\rm e}^{2\xi_{+}}\right)^{-\frac{3}{2}}
+{\rm e}^{-2\xi_{-}}\left(1+{\rm e}^{-2\xi_{-}}\right)^{-\frac{3}{2}}\right]
\label{eq:a1exp}
\end{equation}
for the system of eq.~(\ref{eq:kbark}) in the $\phi^6$ model. It is 
straightforward to compute these coefficients numerically for any prescribed 
value of $X$. This is completely sufficient for the subsequent numerical 
integration of the ODE for $X(t)$ and $A(t)$ that follow from the variational 
principle for the collective coordinate Lagrangians. For the $\varphi^4$ model 
analytic expressions for some of the coefficients have been derived some time 
ago~\cite{Campbell:1983xu,Sugiyama:1979mi}. Unfortunately, the formulas
presented in ref.~\cite{Sugiyama:1979mi} contain some misprints which
propagated through the literature and make the quoted numerical results
for the collective coordinate approach in the $\varphi^4$ model 
unreliable. We therefore revisit those calculations in the next section.
The subsequent sections contain novel studies on the collective coordinate 
approach in the $\phi^6$ model.

\section{Revisiting the $\varphi^4$ model}

The most impressive result from the collective coordinate approach 
in the $\varphi^4$ model is the prediction for the critical velocity
$v_{\rm cr}\approx0.289$~\cite{Goodman:2005aa} above which bounces and 
traps cease to exist. This is only about 10\% off from the exact 
result from the PDE, 0.26~\cite{Campbell:1983xu}. However, a number of 
(uncontrollable) approximations to the system were used: All non--harmonic
terms were omitted ($a_{10},\ldots,a_{13}=0$), the ODE were diagonalized
($a_{6},\ldots,a_{9}=0$) and the direct interactions between the shape modes
at $\pm X(t)$ were discarded via $a_i\,\mapsto\,\lim_{X\to\infty}a_i(X)$ 
for $i=3,4$~\cite{Anninos:1991un}.  The latter approximation also avoids the 
null--vector problem discussed in ref.~\cite{Caputo:1991cv}

\begin{figure}[t]
\centerline{
\includegraphics[width=7.0cm,height=4.3cm]{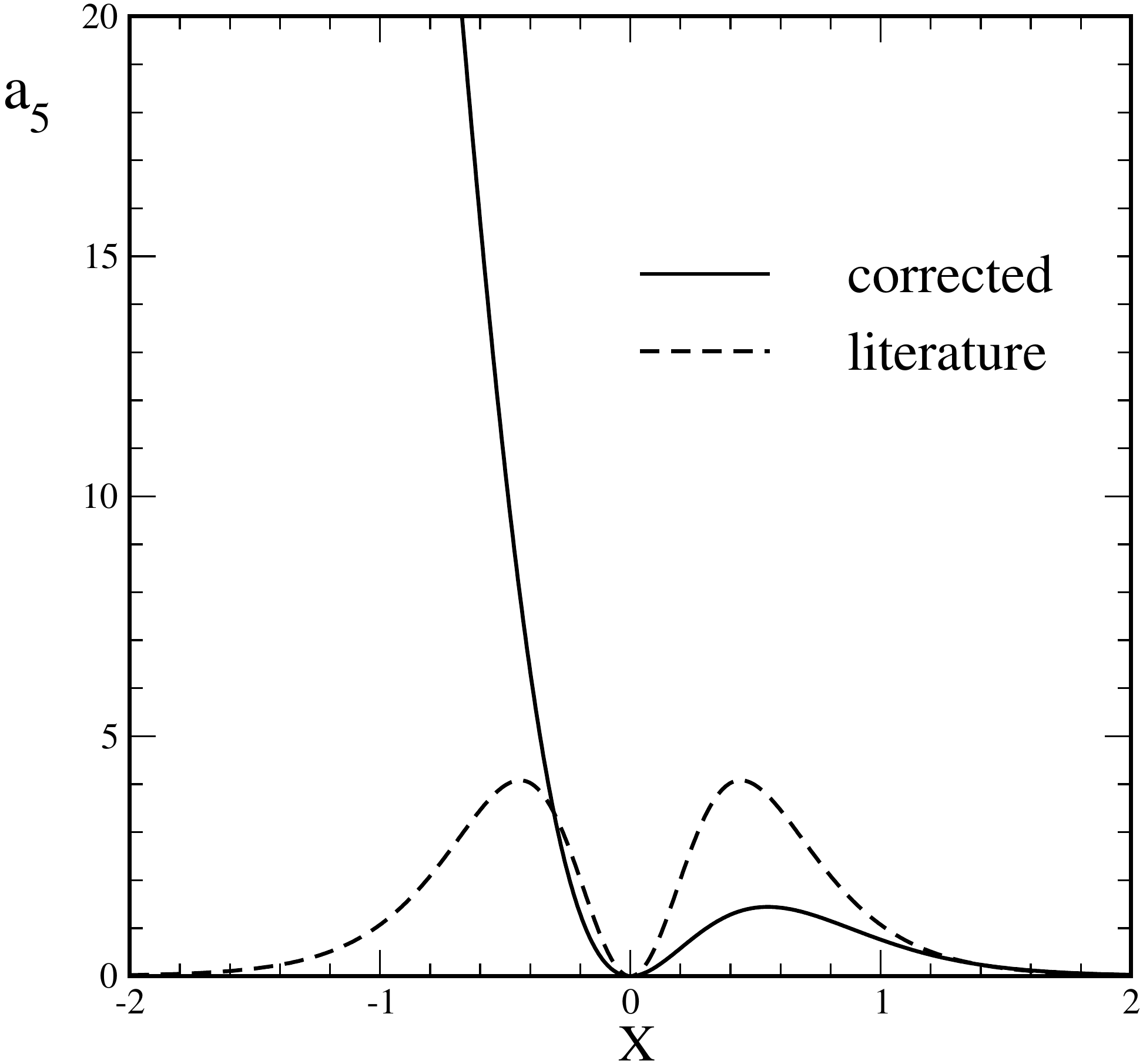}\hspace{1cm}
\includegraphics[width=7.0cm,height=4.3cm]{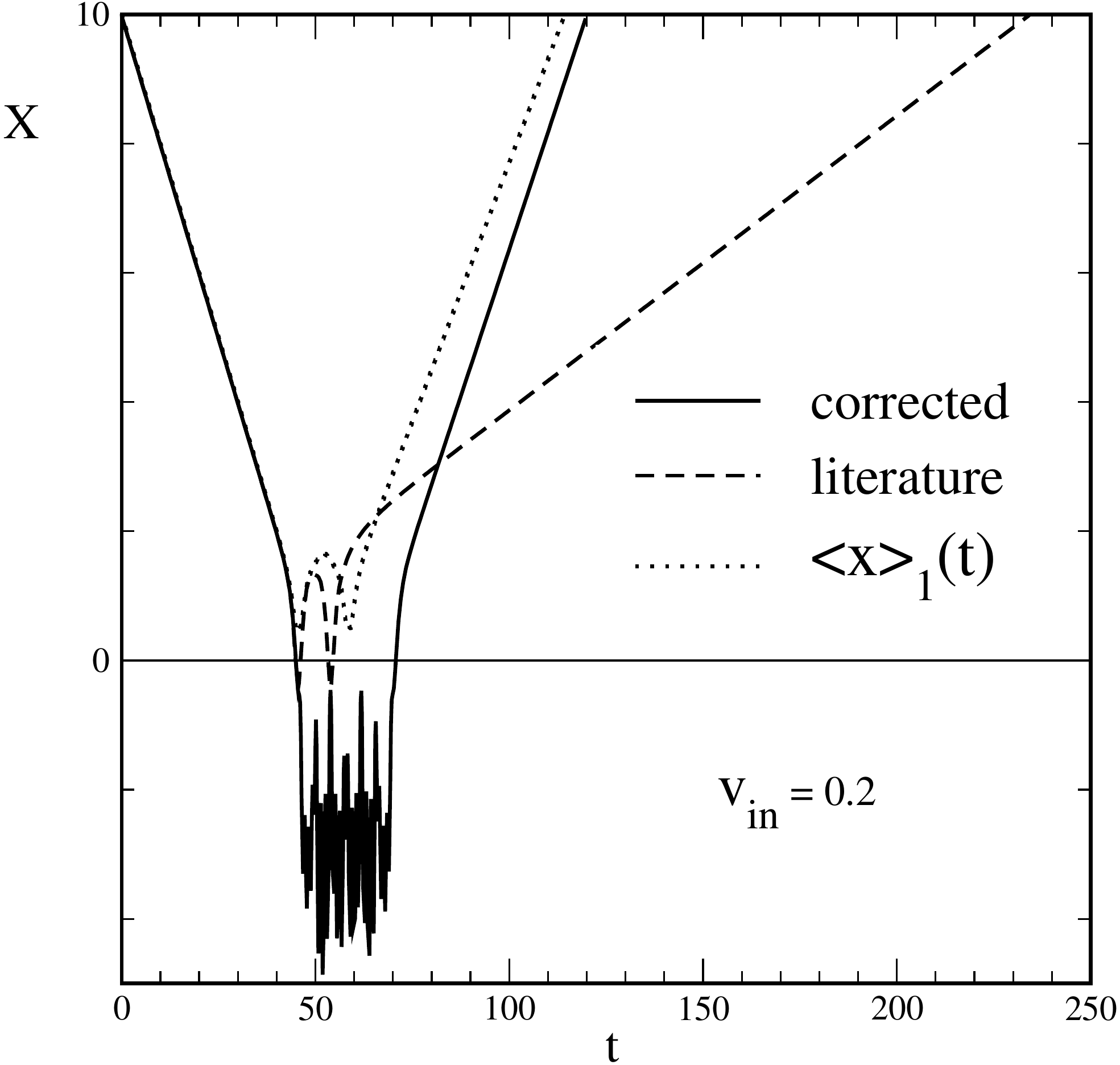}}
\caption{\label{fig:a5}Left panel: Correction of the coefficient function
$a_5(X)$ for $v_{\rm in}=0$ ({\it corrected} refers to 
eq.~(\ref{eq:a5corr}), {\it literature}
denotes the expression derived in ref.~\cite{Sugiyama:1979mi} and frequently
adopted thereafter). Right panel: comparison of the resulting time
dependence of the collective coordinate $X(t)$; the dotted line 
stems from the PDE with $n=1$ in eq.~(\ref{eq:xn}).}
\end{figure}

The source term for the shape mode, $a_5$ is interesting because it 
represents a small amplitude variation about the kink--antikink configuration. 
It should hence vanish only when $\varphi_{K\overline{K}}$ is a static 
solution to the field equation. To gain some insight into the form
of $a_5$ before computing it explicitly it is instructive to consider
\begin{equation}
\varphi_{K\overline{K}}(0,X)=2{\rm tanh}(X)-1\,
\longrightarrow 
\begin{cases}
1 & \mbox{for}\quad X\to+\infty\cr
-3 & \mbox{for}\quad X\to-\infty\,.
\end{cases}
\label{eq:ona5}
\end{equation}
The second case suggests that $\varphi_{K\overline{K}}(x,-\infty)$ is 
not a solution to the field equations because then the center between widely 
separated kink and antikink profiles is not a vacuum configuration. But 
$\varphi(x,+\infty)$ is a valid solution. Hence, $a_5$ cannot be symmetric
under $X\to-X$. However, that symmetry is erroneously reflected by the 
expression in the appendix of ref.~\cite{Sugiyama:1979mi}, wherein it
is called $F(X)$. Unfortunately, this incorrect expression has been used 
in all subsequent studies of the ODE. Redoing the calculation shows that 
the correct result is (for $v_{\rm in}=0$)
\begin{equation}
a_5(X)=3\sqrt{6}\pi\left[ 2-2\,{\rm tanh}^3(X)
-\frac{3}{{\rm cosh}^2(X)}+\frac{1}{{\rm cosh}^4(X)}\right]\,,
\label{eq:a5corr}
\end{equation}
which obviously is not symmetric under $X\to-X$. Changing the power of the 
${\rm tanh}(X)$ term from~3 to~2 and doubling the arguments of all hypergeometric 
functions reproduces the expression in ref.~\cite{Sugiyama:1979mi}.
In figure~\ref{fig:a5} the consequences of correcting this coefficient
function are studied.
As discussed above, the invariance of $a_5$ under reflection is lost. More 
importantly the solutions to the Euler--Lagrange equations for $X(t)$ and 
$A(t)$ that we solve for the initial conditions\footnote{Numerically we 
find that $X(0)\sim10$ is an adequate representation of infinity as no 
overlap between the initial structures occurs.}
\begin{equation}
X(0)\to\infty\,,\quad \dot{X}(0)=\frac{-v_{\rm in}}{\sqrt{1-v_{\rm in}^2}}\,,
\quad A(0)=0\quad {\rm and}\quad \dot{A}(0)=0\,,
\label{eq:ODE0}
\end{equation}
change drastically when corresponding initial conditions and all of 
the above listed approximations are imposed. In figure~\ref{fig:a5} this is 
exemplified for the kink--antikink distance for the initial relative velocity 
$v_{\rm in}=0.2$. Most notably, once the correction is installed, 
multiple bounces occur at any initial velocity.

A main intention is to quantitatively compare the solutions to the ODE 
to those of the PDE for the full field equation. We therefore estimate the 
time--dependent position of the the antikink as the expectation value
\begin{equation}
\langle x \rangle_n(t)=\frac{\int_0^\infty dx\, x\, \epsilon_4^n(t,x)}
{\int_0^\infty dx\,  \epsilon_4^n(t,x)}\,.
\label{eq:xn}
\end{equation}
Here
\begin{equation}
\epsilon_4(t,x)=\frac{1}{2}\left[\ddot{\varphi}+\varphi^{\prime\prime}
+\left(\varphi^2-1\right)^2\right]
\label{eq:edens4}
\end{equation}
is the energy density for the solution $\varphi=\varphi(x,t)$ to the 
PDE~(\ref{eq:eom1}) for initial configurations described by 
equation~(\ref{eq:ansatz}) at $t=0$. The integer $n$ can be freely 
chosen to eventually emphasize effects. Since the energy density is 
typically strongly localized, increasing $n$ turns the distribution 
against which $x$ is tested, into a $\delta$--function.
From the right panel of figure \ref{fig:a5} we also see that 
the corrected ODE result shows almost no resemblance with the PDE
solution, while the literature (though incorrect) one correctly 
reproduces the number of bounces. However, with the correction for 
$a_5$ installed the out--going velocity is well reproduced.

\begin{figure}[t]
~\vskip-0.4cm
\centerline{~~~
\includegraphics[width=8.0cm,height=5.5cm]{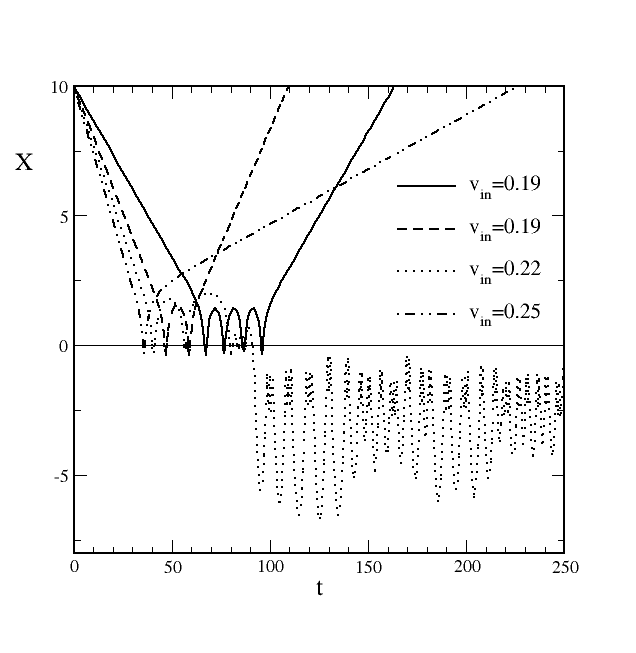}\hspace{0.4cm}
\includegraphics[width=8.0cm,height=5.4cm]{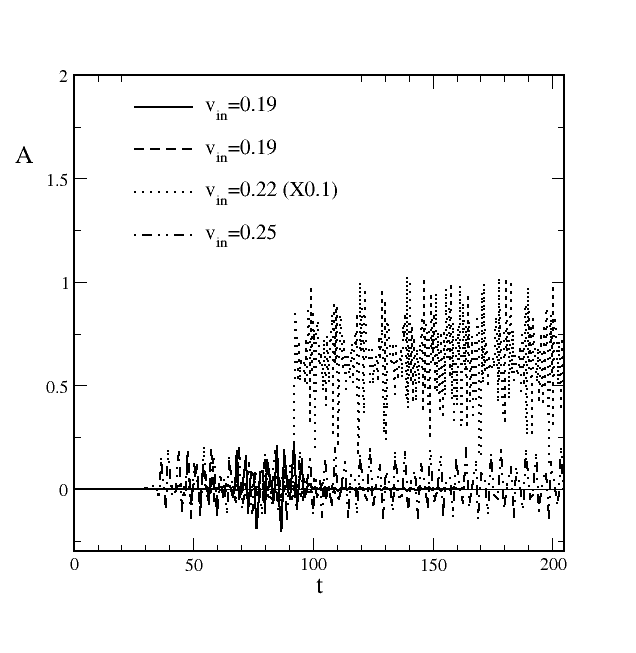}}
\caption{\label{fig:ODEphi4}Solution to the ODE in the 
$\varphi^4$ model. Left panel: kink--antikink separation,
right panel: amplitude of shape mode (note the change of
scale in the $v_{\rm in}=0.22$ entry).}
\end{figure}

In view of these results we must question the above listed approximations.
This is even more the case as the consistency conditions emerging from 
$\varphi_{K\overline{K}}(x,\infty)$ being a solution for $v_{\rm in}\ne0$ 
read\footnote{We have verified them both analytically and numerically.}
\begin{equation}
\lim_{X\to\infty}\,\left[a_5(X)+\,a_6(X)\,\dot{X}^2(0)\right]=0
\qquad {\rm and}\qquad \lim_{X\to\infty}\,a_7(X)=0
\label{eq:consist}
\end{equation}
and thus prohibit the omission of $a_6$. We have numerically 
solved the ODE for $X(t)$ and $A(t)$ with the sole approximation being that
on $a_{3,4}$ to avoid the null--vector problem. (We comment on possible
improvements in the summary section.). The results of these calculations
are shown in figure~\ref{fig:ODEphi4}.
It turns out that with the correction on $a_5$, the corresponding 
contribution to the energy, $E_5=-a_5(X)A$ may absorb much energy
when $X$ becomes negative. Hence trapping type solutions with large
amplitudes of the shape mode emerge. A posteriori, this again
disqualifies the harmonic approximation. The entry
with $v_{\rm in}=0.22$ in figure \ref{fig:ODEphi4}
is a typical example thereof. Most notably, however, we find that
without the many approximations, the ODE indeed predict a critical 
velocity of $v_{\rm cr}=0.247$ above which trapping or bounce type
solution cease to exist. This compares favorably to the PDE result 
of~$0.26$~\cite{Campbell:1983xu}.

\section{Antikink--Kink in $\phi^6$}

In the $\phi^6$ model there are two types of initial structures that 
contain kink and antikink configurations. We first consider the so--called
{\it antikink--kink} ansatz~\cite{Dorey:2011yw} of eq.~(\ref{eq:kbark}) 
that, as in the $\varphi^4$ model, eq.~(\ref{eq:kkbar4}), parameterizes 
a legitimate solution only for $X\to\infty$ but not for $X\to-\infty$. 
Hence it cannot describe penetration either. 
To compare the solutions to the ODE to those of the PDE for the 
full field equation in the $\phi^6$ model we need to replace the
energy density in eq.~(\ref{eq:xn}) by
\begin{equation}
\epsilon_6(t,x)=\frac{1}{2}\left[\ddot{\phi}+\phi^{\prime\prime}
+\phi^2\left(\phi^2-1\right)^2\right]\,.
\label{eq:edens6}
\end{equation}
Here $\phi=\phi(x,t)$ are the solutions to the PDE~(\ref{eq:eom2}) with 
initial conditions taken as the $t=0$ configuration in eq.~(\ref{eq:kbark}).

\begin{figure}
\centerline{~
\includegraphics[width=6.8cm,height=4.1cm]{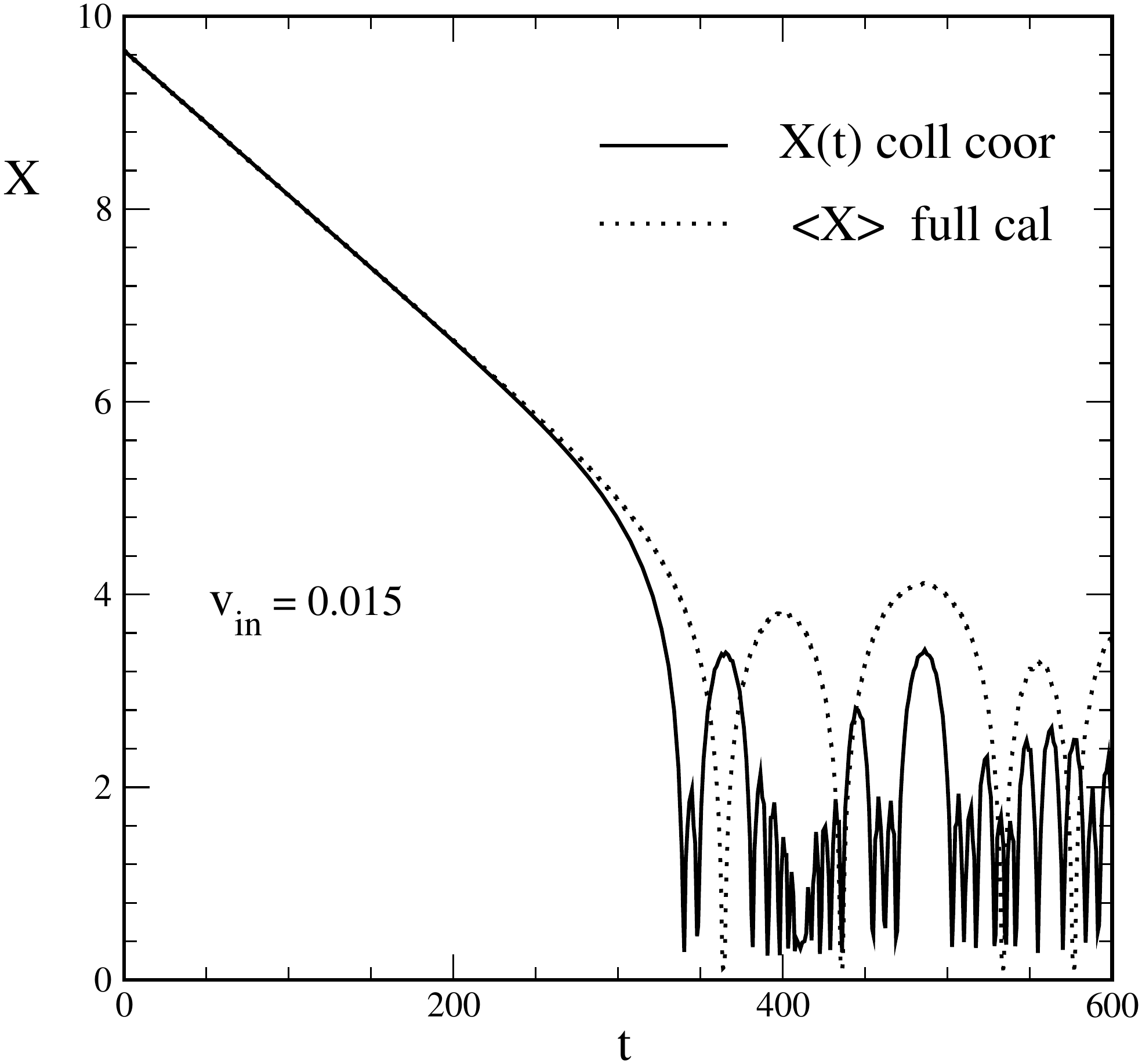}\hspace{1cm}
\includegraphics[width=7.0cm,height=4.1cm]{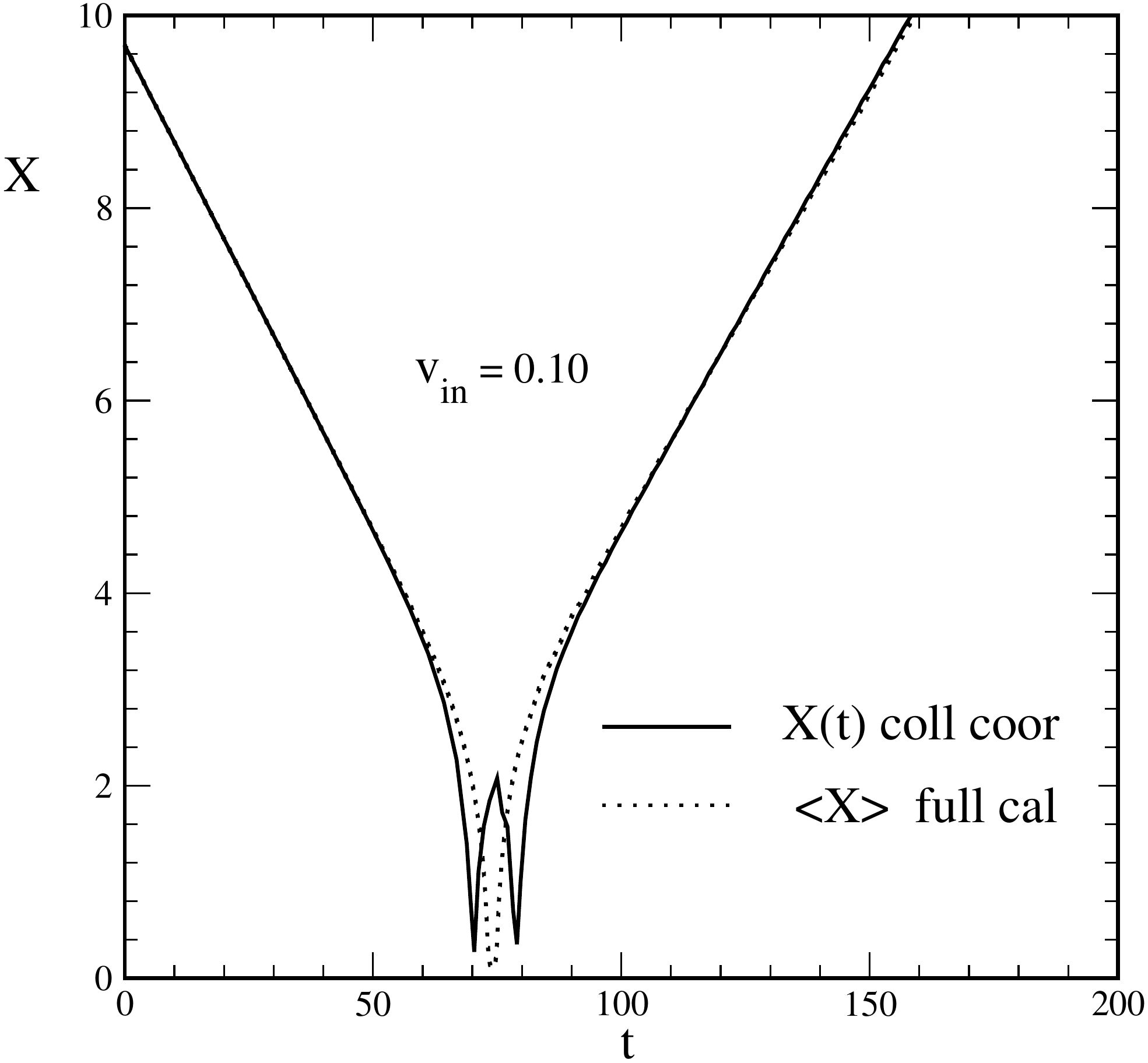}}
\bigskip

\centerline{
\includegraphics[width=6.8cm,height=4.1cm]{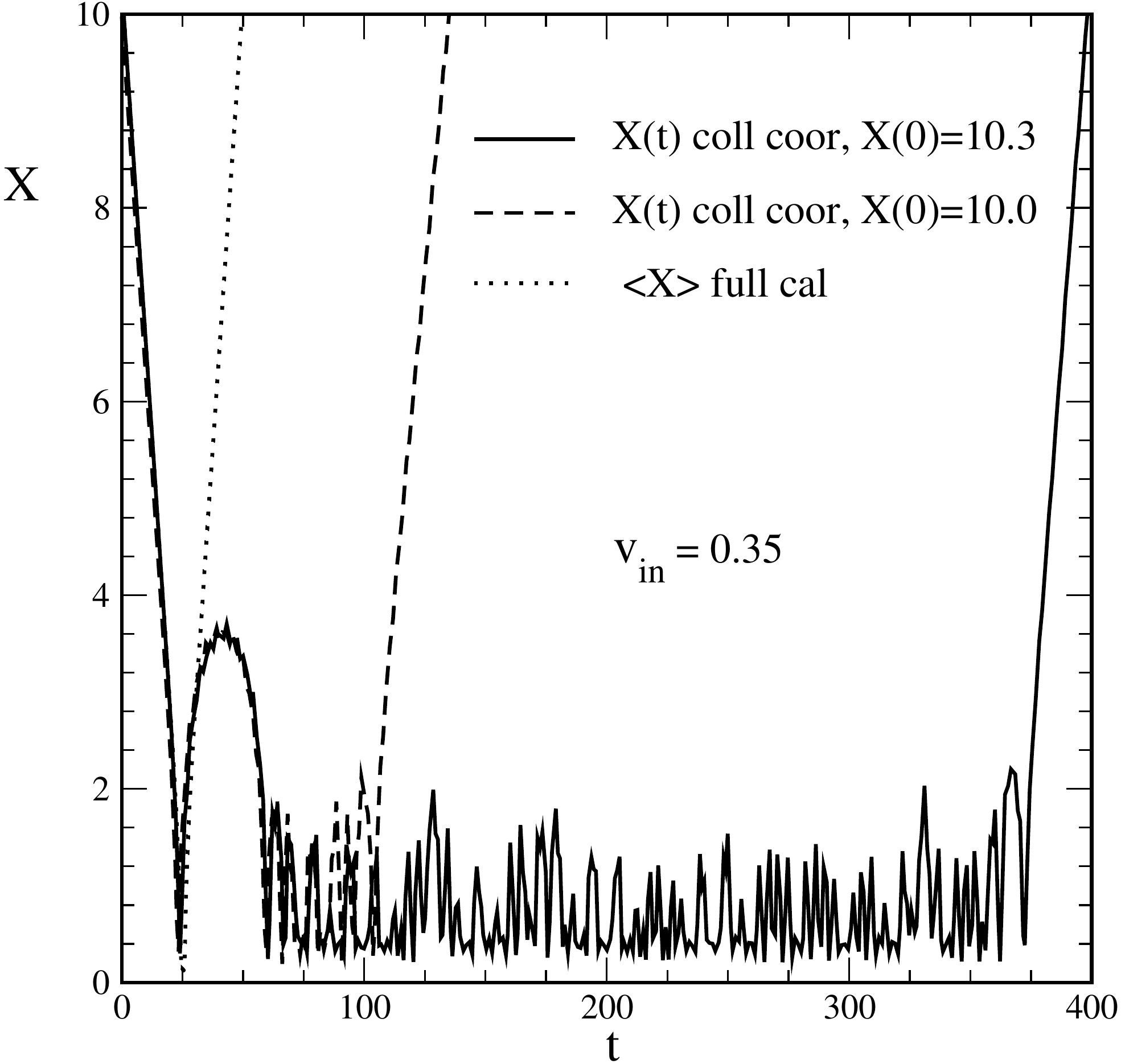}\hspace{1cm}
\includegraphics[width=6.8cm,height=4.1cm]{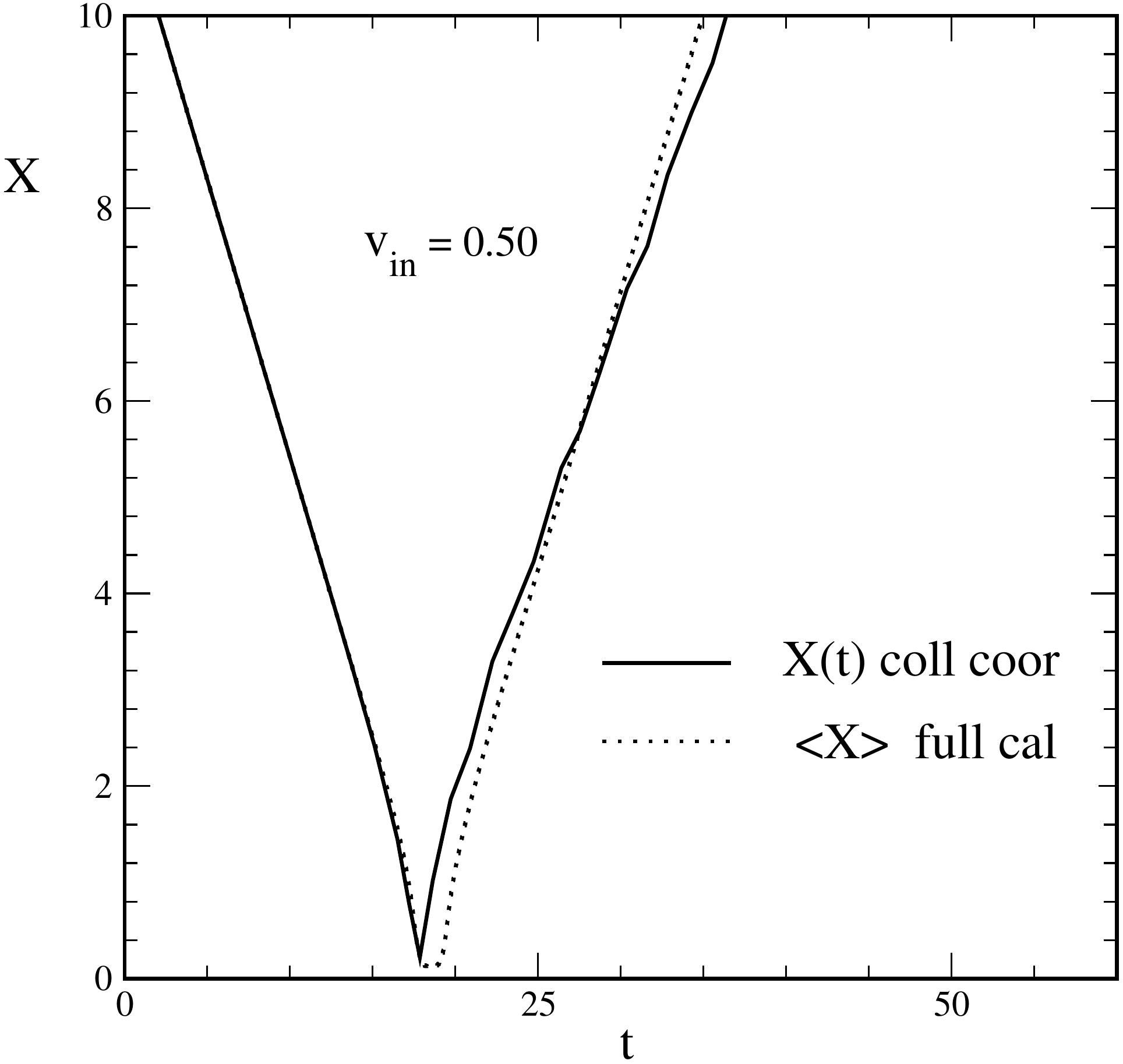}}
\caption{\label{fig:kbark6}Solutions to the ODE for 
antikink--kink scattering in the $\phi^6$ model and comparison to 
PDE results ({\it full cal}) with $n=1$ in eqs.~(\ref{eq:xn}) 
and~(\ref{eq:edens6}).}
\end{figure}

The PDE yields a critical velocity of $v_{\rm cr,PDE}=0.05$ above which 
no traps or bounces occur. As shown in figure \ref{fig:kbark6} the solutions 
to the ODE and PDE differ substantially at low initial velocities. Typically 
the bounce frequency from the ODE is much larger. So is the predicted 
critical velocity $v_{\rm cr,ODE}=0.357$.  It is thus suggestive that 
the collective coordinate approach over--estimates the attraction between 
kink and antikink. For velocities above $v_{\rm cr,ODE}$ this can also be 
observed as the ODE prediction for the final velocity is lower than that 
from the PDE. In the close vicinity of that critical velocity the technical 
parameters that enter the numerical treatment of the ODE seem to matter, 
thereby indicating chaotic behavior. In particular, the results are 
sensitive to the coordinate value that is supposed to represent infinity; 
as can be seen from the $v_{\rm in}=0.35$ entry.

\begin{figure}
\centerline{
\includegraphics[width=7.1cm,height=4.1cm]{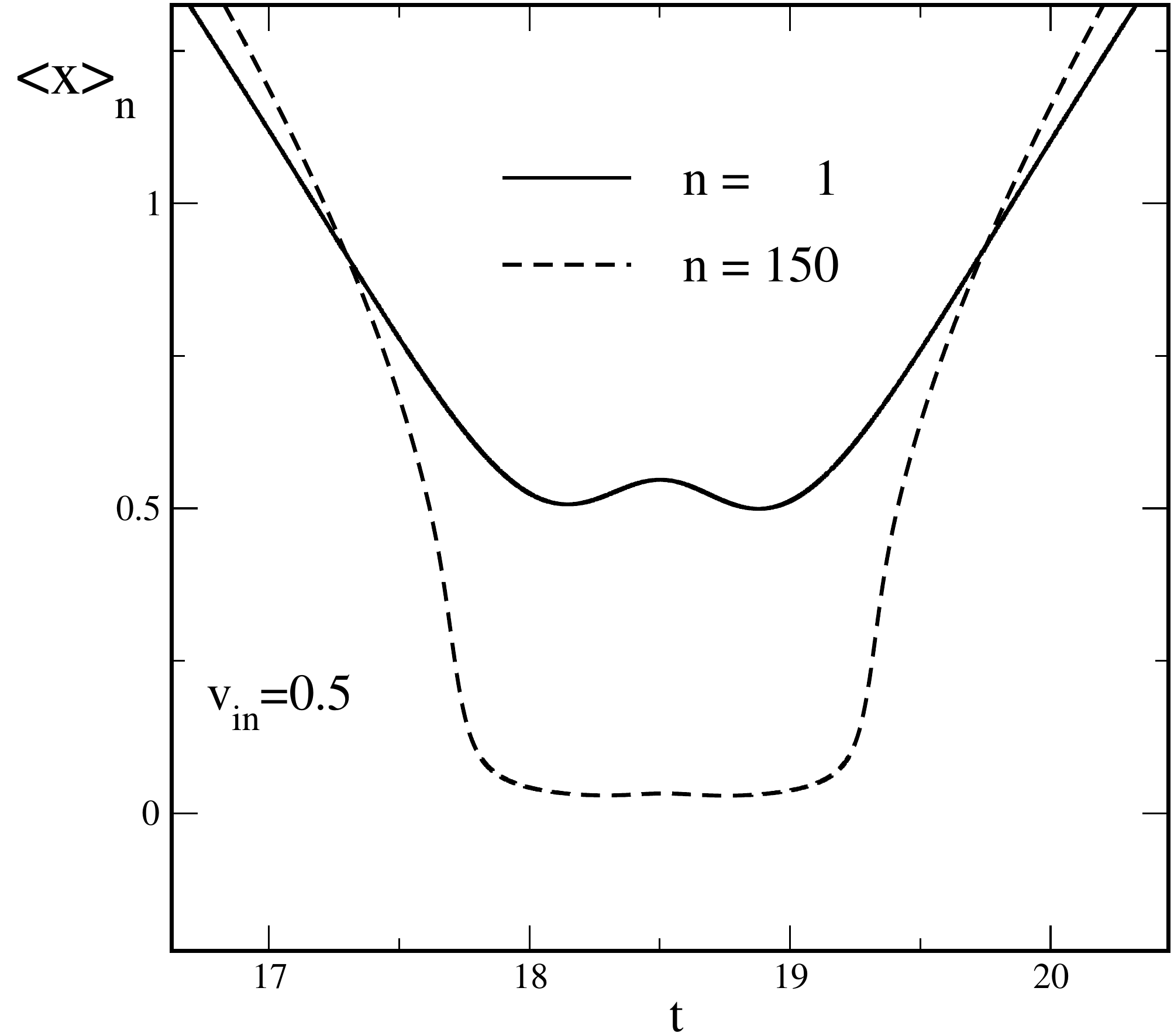}\hspace{0.9cm}
\includegraphics[width=7.1cm,height=4.1cm]{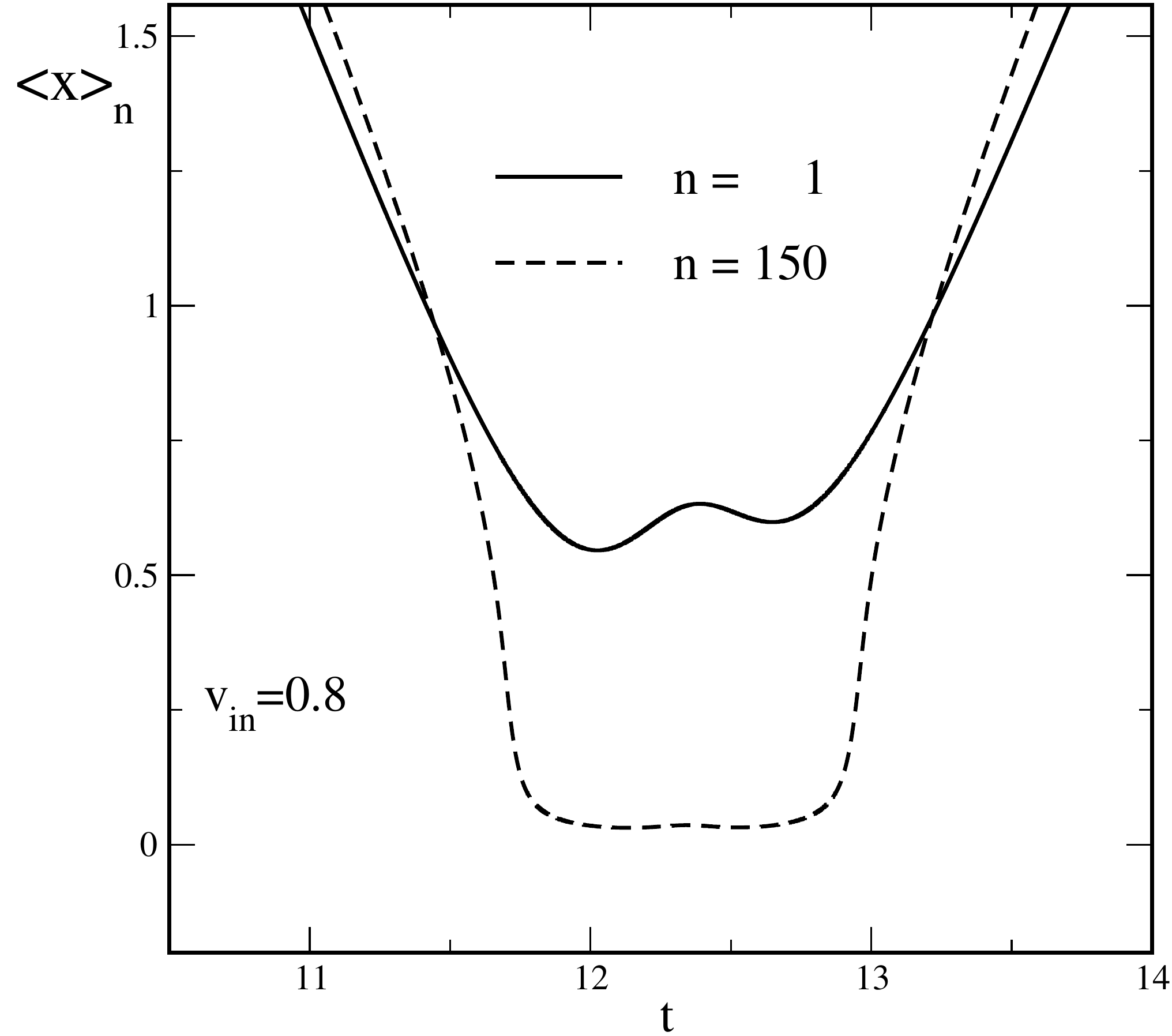}~~}
\caption{\label{fig:minib}Pseudo--bounces in the antikink--kink system
of the $\phi^6$ model from the PDE . The parameter
$n$ enters the computation the expectation value of the position of
the antikink via eq.~(\ref{eq:xn}).}
\end{figure}

For large initial velocities the PDE results actually exhibit some
pseudo--bounces shown in figure~\ref{fig:minib}. The kink and antikink 
linger on top of each other for a moderate time interval. If the distribution
in eq.~(\ref{eq:xn}) is taken to be broad, the trajectory feigns mini--bounces.

\section{Kink--Antikink in $\phi^6$}

\begin{figure}
\centerline{
\includegraphics[width=6.8cm,height=4.5cm]{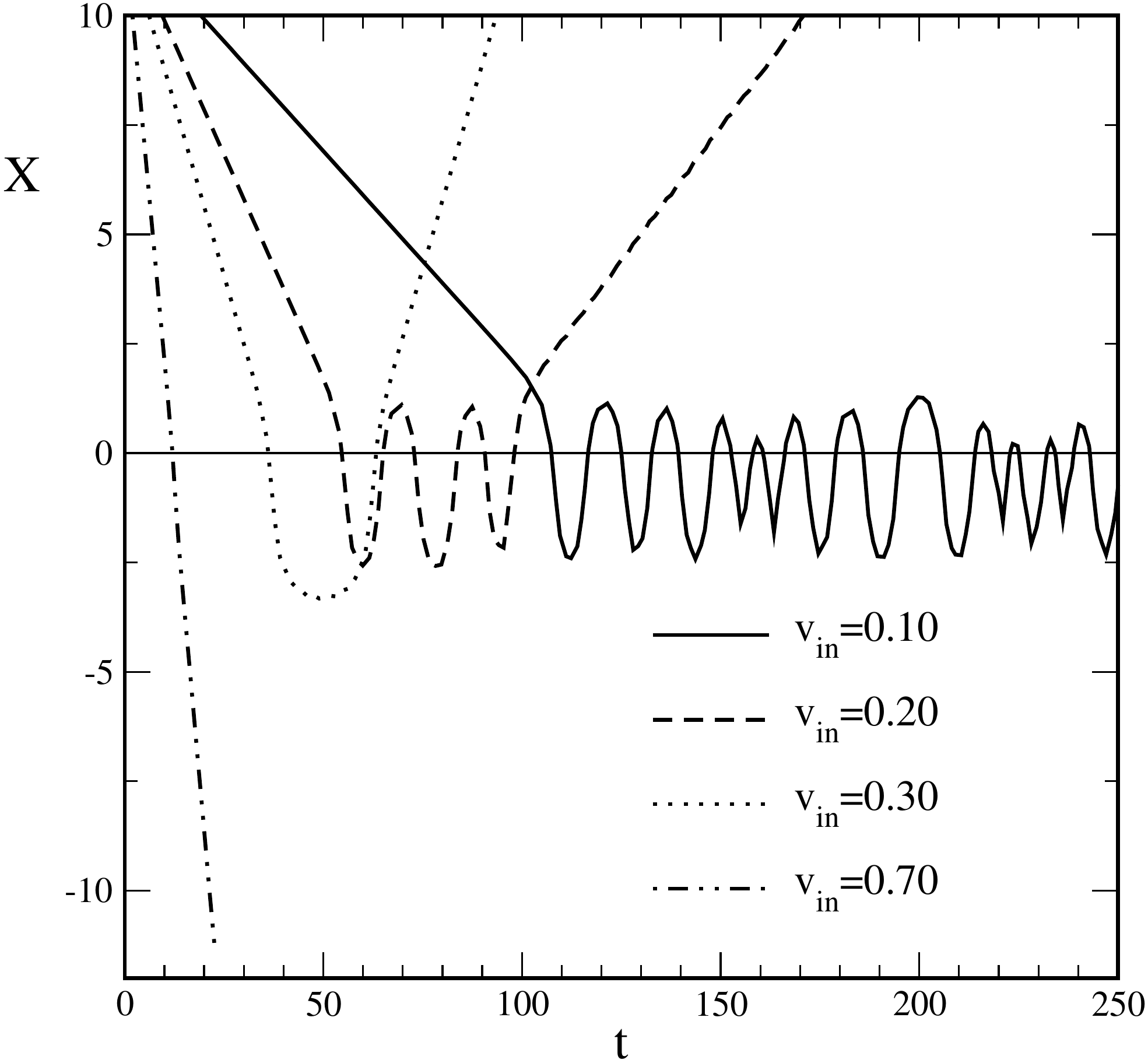}\hspace{1cm}
\includegraphics[width=6.8cm,height=4.5cm]{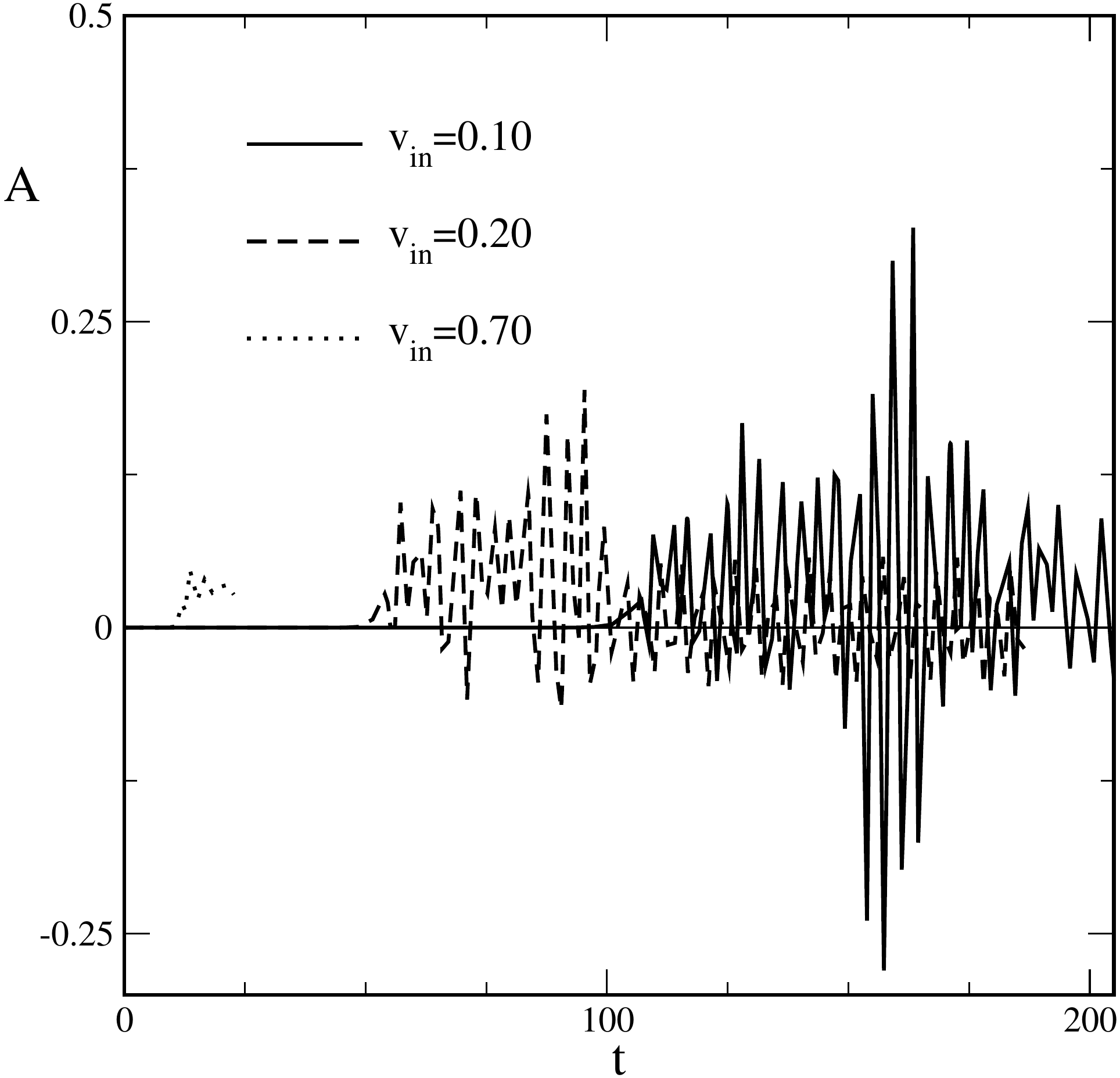}}
\caption{\label{fig:kkbar0}Solutions for kink--antikink interactions 
from the ODE system in the $\phi^6$ model. Left panel: kink--antikink 
separation, right panel: amplitude of shape mode.}
\end{figure}

In contrast to the previously discussed configurations the 
so--called {\it kink--antikink} structure from eq.~(\ref{eq:kkbar}) 
has topological properties which would allow it to be a solution 
even in the case $X\to-\infty$. However, the potential $a_2$ 
in eq.~(\ref{eq:lagf}) is not symmetric under $X\to-X$ and
$a_2(-\infty)>a_2(\infty)$. Hence there is a velocity threshold 
for solutions that correspond to the kink penetrating the antikink.
Simple kinematical considerations on the ODE coefficients $a_i$ 
indicate this threshold to be $v_{\rm th}=0.296$. The numerical 
solution to the ODE yields a slightly larger value,~0.302. The 
left panel in figure~\ref{fig:kkbar0} shows that the configuration 
with initial velocity $v_{\rm in}=0.30$ bounces while bigger values 
of $v_{\rm in}$ yield $X(t)\to-\infty$ at large times.
The ODE yields a window $0.216\le v_{\rm in}\le 0.302$ in which the
system bounces exactly twice. As for the $\varphi^4$ model, multiple 
bounce solutions are characterized by a sizable amplitude of the shape 
mode.  Even though this is not a valid vibrational solution it 
supports the working hypothesis that, once the collective coordinate
ansatz allows for its excitation, it will do so. It should also be
noted that the upper bound of the above mentioned window agrees 
reasonably well (but not perfectly) with the critical velocity (0.289) 
from the PDE calculation. 

\begin{figure}
\centerline{
\includegraphics[width=6.8cm,height=4.5cm]{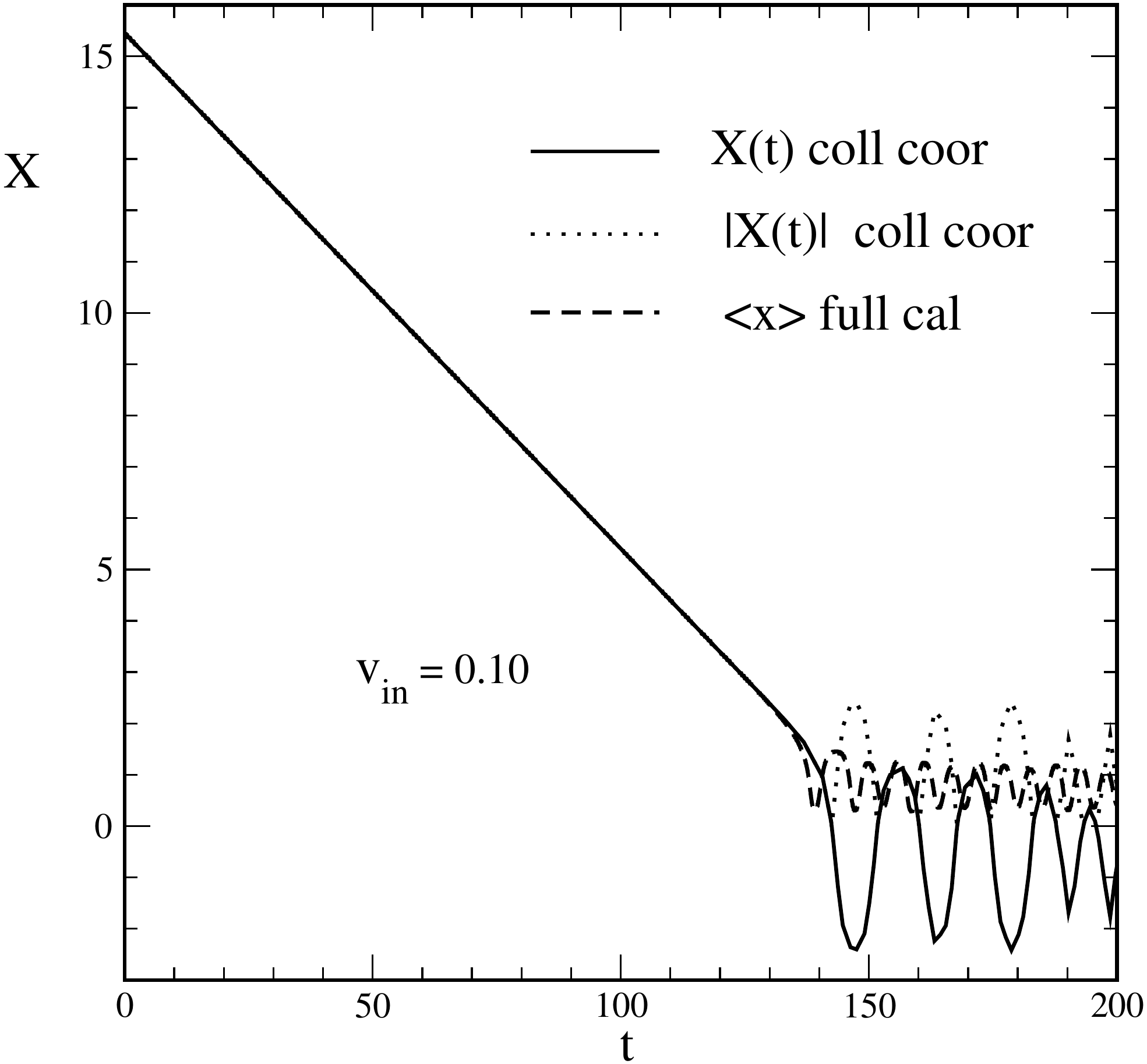}\hspace{1cm}
\includegraphics[width=6.8cm,height=4.5cm]{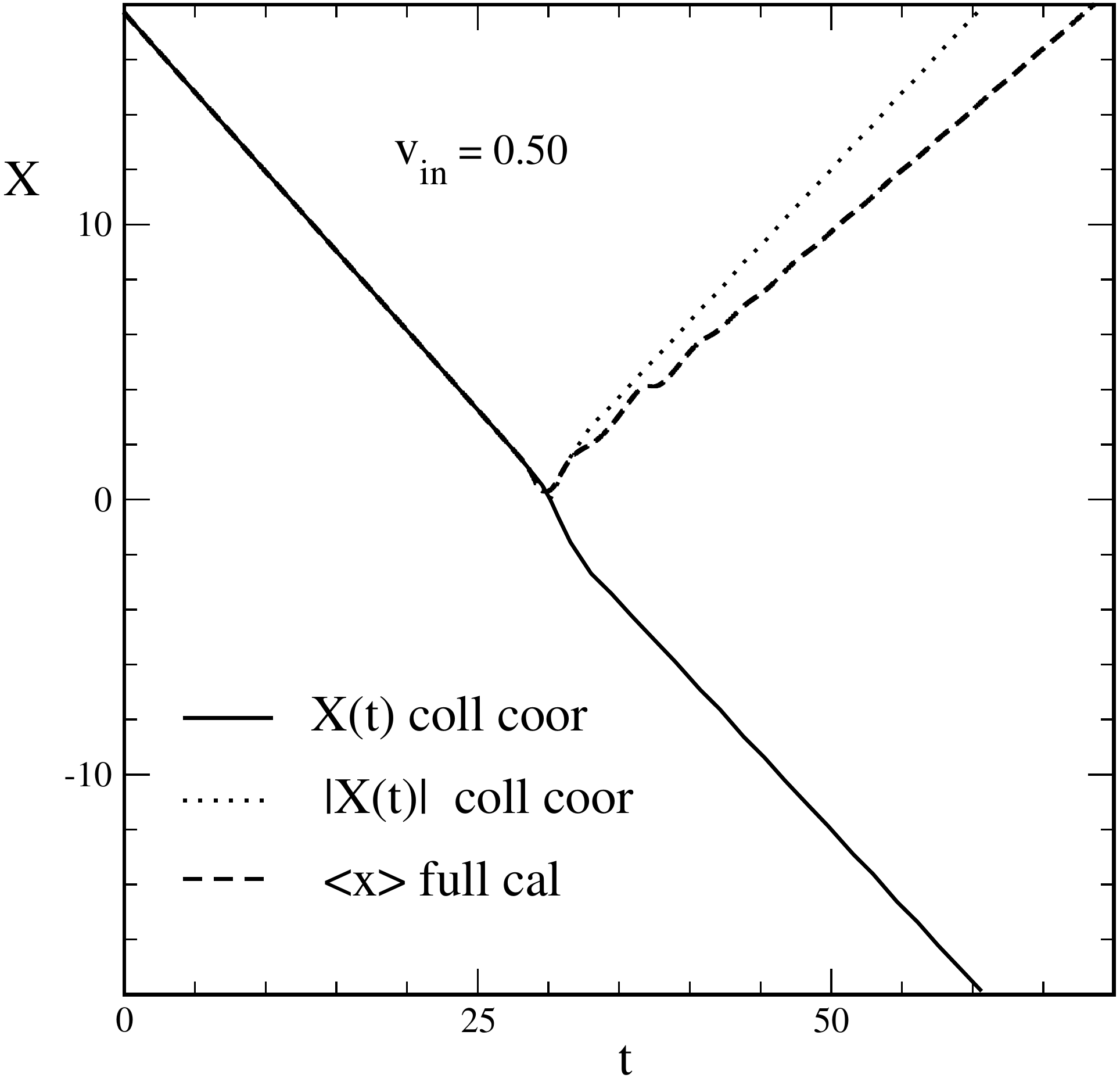}}
\caption{\label{fig:kkbar1}Comparison of PDE and ODE solutions for
kink--antikink interactions in the $\phi^6$ model. {\it Full cal} 
refers to $\langle x\rangle_1$ in eqs.~(\ref{eq:xn}) 
and~(\ref{eq:edens6}).}
\end{figure}

For $v_{\rm in}=0.10$ and $v_{\rm in}=0.50$ we compare the PDE and ODE 
results for the distances between kink and antikink figure~\ref{fig:kkbar1}. 
The PDE calculation for $\langle x\rangle$ is as in 
eqs.~(\ref{eq:eom2}),~(\ref{eq:xn}) and~(\ref{eq:edens6}). However, the 
initial condition at $t=0$ is now taken from eq.~(\ref{eq:kkbar}). The 
structure within ODE and PDE solutions agree. However, if applicable, the 
frequency of bounces differs considerably.

\section{Summary}

Collective coordinate calculations are quite useful in non--linear 
field  theory not only because they simplify the field equations 
considerably but also because they potentially provide insight into 
the underlying dynamics by focusing on particular modes. Within the 
$\varphi^4$ and $\phi^6$ models in one time and one space dimension
we have compared the ODE solutions from the collective coordinate 
approaches to those from the PDE for the space--time dependent fields.
Our analysis, however, has revealed that care in needed when attempting 
to draw rigorous conclusions from collective coordinate calculations. 
Not only do they depend on the specific ans\"atze but also on the 
approximations made. For example, the harmonic approximation in the 
$\varphi^4$ model incorrectly yields solutions with repeated bounces 
for any relative initial velocity of the kink--antikink configuration 
(after correcting a literature error within the collective coordinate 
Lagrangian). On the other hand the PDE yields quite a small critical 
velocity beyond which multiple bounces cease to exist.

Our studies support the working hypothesis that a shape mode in the 
vibration spectrum of the kink is not a necessity for bounces and traps 
to occur in kink--antikink scattering. We started from the conjecture that 
a collective coordinate ansatz including such a mode would represent the 
full field configuration reasonably well regardless of whether or not it is 
a true bound state of the kink. We have explored the (dis)agreements
between the ODE and PDE results. These (dis)agreements are qualitatively 
equal in the $\varphi^4$ and $\phi^6$ models. Since the shape mode only 
exists within the $\varphi^4$ model the collective coordinate approach 
can therefore not be used to establish that the shape mode is responsible 
for the existence of multiple bounces in kink--antikink scattering.

Though we have successfully waved a number of previous approximations within 
the ODE calculation, we have still omitted the interaction between the shape 
modes at $\pm X$. Otherwise the calculations run into the null--vector
problem~\cite{Caputo:1991cv}. This problem arises because by pure 
construction of the ansatz, the coefficient of $A$ vanishes for $X=0$,
{\it cf.} eq.~(\ref{eq:ansatz}). The particular combination of 
shape modes at $\pm X$ is commonly assumed because the more general
parameterization with three collective coordinates ($X,A,B$)
\begin{equation}
\varphi_{\rm cc}(x,t)=\varphi_{K\overline{K}}(x,X(t))+
\sfract{3}{2}\left[A(t)\chi_1(x+X(t))+B(t)\chi_1(x-X(t))\right]
\label{eq:ansatz2}
\end{equation}
only yields a linear source term for $A-B$. Of course, the combination
$A+B$ will nevertheless be excited by non--linear and/or higher order effects. 
It will therefore be interesting to investigate the above parameterization.
This will complicate the collective coordinate approach slightly 
as it turns it into a $3\times3$ problem. Beyond that, the ODE approach
is probably no longer a sensible simplification to the PDE.

\section*{Acknowledgments}
The author is grateful to A.~M.~H.~H.~Abdelhady for his collaboration 
at early stages of this study and his comments on the manuscript.
The organizers of the conference are thanked 
for providing the opportunity to present these results. Support from 
the National Research Foundation of South Africa is acknowledged.


\section*{References}
\begin{thebibliography}{9}
\bibitem{Ra82}
R.~Rajaraman, {\em Solitons and Instantons},
\newblock North Holland, 1982.

%\cite{Vilenkin:1994}
\bibitem{Vilenkin:1994}
  A.~Vilenkin and E.P.S.~Shellard,
  \textsl{Cosmic Strings and other Topological Defects},
  Cambridge University Press, Cambridge (UK), (1994).

%\cite{Vachaspati:2006}
\bibitem{Vachaspati:2006}
  T.~Vachaspati, \textsl{Kinks and domain walls: An introduction
  to classical and quantum solitons},
  Cambridge University Press, Cambridge (UK), (2006).

%\cite{Manton:2004}
\bibitem{Manton:2004}
  N.~Manton and P.~Sutcliffe, \textsl{Topological Solitons},
  Cambridge University Press, Cambridge, (UK), (2004).

%\cite{Weigel:2008zz}
\bibitem{Weigel:2008zz}
  H.~Weigel,
  \textsl{Chiral Soliton Models for Baryons},
  Lect.\ Notes Phys.\  {\bf 743} (2008) 1.
  %%CITATION = LNPHA,743,1;%%

%\cite{Bishop:1978}
\bibitem{Bishop:1978}
  A.~R. Bishop and T.~Schneider (Eds.),
  \textsl{Solitons in Condensed Matter Physics},
  Springer Verlag, Berlin (1978).

\bibitem{Abdelhady}
  A.~M.~H.~H.~Abdelhady, {\it Scattering in Soliton Models and Crossing 
   Symmetry}, MSc Thesis, Stellenbosch University (2012), unpublished.

\bibitem{Ra82_5}
  See, for example, chapter 5 in ref.~\cite{Ra82}.

\bibitem{Goodman:2005aa}
  R.~H.~Goodman and R.~Haberman,
  %``Kink-Antikink collisions in the phi**4 equation: the
  % n-bounce resonance and the separatrix map,''
  SIAM J.\ App.\ Dyn.\ Sys.\ {\bf 4} (2005) 1195.

%\cite{Dorey:2011yw}
\bibitem{Dorey:2011yw}
  P.~Dorey, K.~Mersh, T.~Romanczukiewicz and Y.~Shnir,
  %``Kink-antikink collisions in the phi^6 model,''
  Phys.\ Rev.\ Lett.\  {\bf 107} (2011) 091602.
  %[arXiv:1101.5951 [hep-th]].

%\cite{Sugiyama:1979mi}
\bibitem{Sugiyama:1979mi}
  T.~Sugiyama,
  %``Kink - Antikink Collisions In The Two-dimensional Phi**4 Model,''
  Prog.\ Theor.\ Phys.\  {\bf 61} (1979) 1550.
  %%CITATION = PTPKA,61,1550;%%

%\cite{Abdelhady:2011tm}
\bibitem{Abdelhady:2011tm}
  A.~M.~H.~H.~Abdelhady and H.~Weigel,
  %``Wave-Packet Scattering off the Kink-Solution,''
  Int.\ J.\ Mod.\ Phys.\ A {\bf 26} (2011) 3625.
  %%CITATION = ARXIV:1106.3497;%%

%\cite{Kudryavtsev:1975dj}
\bibitem{Kudryavtsev:1975dj}
  A.~E.~Kudryavtsev,
  %``About Soliton Similar Solutions for Higgs Scalar Field,''
  Pisma Zh.\ Eksp.\ Teor.\ Fiz.\  {\bf 22} (1975) 178.
  %%CITATION = ZFPRA,22,178;%%

\bibitem{Campbell:1983xu}
  D.~K.~Campbell, J.~F.~Schonfeld and C.~A.~Wingate,
  %``Resonance Structure in Kink - Antikink Interactions in $\phi^{4}$ Theory,''
  Physica {\bf 9D} (1983) 1.
  %%CITATION = PHYSA,9D,1;%%

%\cite{Belova:1985fg}
\bibitem{Belova:1985fg}
  T.~I.~Belova and A.~E.~Kudryavtsev,
  %``Quasiperiodical Orbits In The Scalar Classical Lambda Phi**4 Field Theory,''
  Physica D {\bf 32} (1988) 18.
  %%CITATION = PHYSA,D32,18;%%

%\cite{Anninos:1991un}
\bibitem{Anninos:1991un}
  P.~Anninos, S.~Oliveira and R.~A.~Matzner,
  %``Fractal structure in the scalar lambda (phi**2-1)**2 theory,''
  Phys.\ Rev.\ D {\bf 44} (1991) 1147.
  %%CITATION = PHRVA,D44,1147;%%

\bibitem{Goodman:2004aa}
  R.~H.~Goodman and R.~Haberman,
  % Interaction of Sine-Gordon Kinks with Defects:...
  Physica D {\bf 195} (2004) 303.

%\cite{Caputo:1991cv}
\bibitem{Caputo:1991cv}
  J.~G.~Caputo and N.~Flytzanis,
  %``Kink antikink collisions in sine-Gordon and phi**4 models:
     %Problems in the variational approach,''
  Phys.\ Rev.\ A {\bf 44} (1991) 6219.
  %%CITATION = PHRVA,A44,6219;%%

\end{thebibliography}
\end{document}